\documentclass[a4paper]{PoS}

\usepackage{amsmath}
\usepackage{amssymb}
\usepackage{graphicx}
\usepackage[hyphens]{url}
\usepackage{cite}
\usepackage{array}
\usepackage{multirow}
\usepackage{caption}
\usepackage{subcaption}
\usepackage{mdframed}
\usepackage{color}
\usepackage[usenames,dvipsnames]{xcolor}
\usepackage{listings}
\usepackage{bbold}
\usepackage{cleveref}
\usepackage{verbatim}

\newcommand{\Tr}{\mathrm{Tr}\;}

\newcommand{\SU}[1]{$\mathrm{SU}(#1)$}
\newcommand{\UA}[1]{$\mathrm{U}(#1)_\mathrm{A}$}
\newcommand{\UAa}[1]{$\mathrm{U}(#1)_\mathrm{A}^\mathrm{anom}$}

\graphicspath{{./figs/}}

\title{Chiral Ward identities for Dirac eigenmodes with staggered fermions}

\ShortTitle{Chiral Ward identities for staggered fermions}

\author{\speaker{Hwancheol Jeong}, Sunghee Kim, Weonjong Lee, and
  Jeonghwan Pak \\
  Lattice Gauge Theory Research Center, CTP, and FPRD,\\
  Department of Physics and Astronomy,\\
  Seoul National University, Seoul 08826, South Korea\\
  E-mail: \email{sonchac@gmail.com}, \email{wlee@snu.ac.kr}
  }

\author{Chulwoo Jung\\
  Physics Department, Brookhaven National Laboratory, Upton,
  NY 11973, USA\\
  E-mail: \email{chulwoo@bnl.gov}
  }

\author{SWME Collaboration}

\abstract{
We study chiral properties of eigenvalue spectrum for staggered
quarks.
We present a new method to identify would-be zero modes and nonzero
modes using their symmetry and chiral properties.
Here, we review the traditional method with HYP improved staggered
quarks, and extend it to a completely new method which uses the chiral
Ward identities and leakage patterns to achieve the goal.
}

\FullConference{37th International Symposium on Lattice Field
  Theory - Lattice2019\\
  16-22 June 2019\\
  Wuhan, China}

\begin{document}


\section{Introduction}
\label{sec:intro}
Here, we present recent progress in understanding chiral properties of
staggered Dirac eigenmodes based on our previous works in Refs.~\cite{
  Jeong:2017kst, Cundy:2016tmw}.
%


\section{Eigenvalues of Dirac operators with staggered fermions}
\label{sec:eig}

Let us consider Dirac operator $D_s$ for staggered fermions.
Since $D_s$ is anti-Hermitian, eigenvalues of $D_s$ are purely
imaginary or zero:
\begin{equation}
  \label{eq:ev_eq_d}
  D_s | f_\lambda \rangle = i \lambda\, | f_\lambda \rangle \,,
\end{equation}
where $\lambda$ is real, and $| f_\lambda \rangle$ is an eigenvector
with its eigenvalue $i \lambda$.
$D_s$ also anti-commutes with the operator $\Gamma_\varepsilon =
[\gamma_5 \otimes \xi_5]$ which is the generator for
\UA{1} symmetry.
Here, we adopt the same notation as in Ref.~\cite{Jeong:2017kst}.
Since $\Gamma_\varepsilon$ anti-commutes with $D_s$, one can show that
an eigenstate $| f_\lambda \rangle$ with $\lambda \ne 0$ has its
partner state $| f_{-\lambda} \rangle$ with eigenvalue $-i
\lambda$ \cite{Jeong:2017kst}.
The partner eigenvector $| f_{-\lambda} \rangle$ can be obtained by
applying $\Gamma_\varepsilon$ to the eigenvector $| f_{+\lambda} \rangle$
with phase difference:
\begin{align}
  \label{eq:fp_to_fm}
  \Gamma_\varepsilon\, | f_{+\lambda} \rangle &= e^{+i\theta}\, |
  f_{-\lambda} \rangle \,,
  \qquad
  \Gamma_\varepsilon\, | f_{-\lambda} \rangle = e^{-i\theta}\, |
  f_{+\lambda} \rangle \,.
\end{align}
Here, the phase $\theta$ is real, and also turns out to be arbitrary
\cite{Jeong:2017kst}.
In practice, we do not calculate eigenvalues of $D_s$ directly.
Instead, we use a Hermitian and positive semi-definite operator
$D_s^\dagger D_s$, which satisfies
\begin{equation}
  \label{eq:ev_eq_ddagd}
  D_s^\dagger D_s\, |\, g_{\lambda^2} \rangle = \lambda^2\, |\,
  g_{\lambda^2} \rangle \,.
\end{equation}
%
%
The corresponding eigenvectors $|\, f_{\pm\lambda} \rangle$ are
obtained by decomposing the eigenvector $|\, g_{\lambda^2} \rangle$
using the projection operators as in Ref.~\cite{Jeong:2017kst}.
Since $D_s^\dagger D_s$ is Hermitian, one can make use of Lanczos
algorithm \cite{Lanczos:1950zz} to calculate its eigenvalues and
eigenvectors.
Here, we use the implicitly restarted Lanczos \cite{
  Lehoucq96deflationtechniques} with acceleration by Chebyshev
polynomial \cite{MR736453}.

\begin{table}[t!]
  \small
  \centering
  \vspace*{-5mm}
  \begin{tabular}{>{\raggedleft\arraybackslash}m{0.25\linewidth}
      |>{\raggedright\arraybackslash}m{0.35\linewidth}}
    \hline\hline
    parameter & value \\
    \hline
    gluon action & tree level Symanzik
    \cite{Luscher:1984xn,Luscher:1985zq,Alford:1995hw} \\
    tadpole improvement & yes \\
    $\beta$ & 5.0 \\
    geometry & $20^4$ \\
    $a$ & 0.077 fm \\
    \hline
    valence quarks & HYP staggered fermions
    \cite{Lee:2002ui,Kim:2010fj,Kim:2011pz} \\
    $N_f$ & $0$ (quenched QCD) \\
    \hline\hline
  \end{tabular}
  \caption{Input parameters for the numerical study. For more details,
    refer to Ref.~\cite{Follana:2005km}.}
  \label{tab:gconf}
\end{table}
All the numerical calculations are performed on the gauge
ensemble described in Table~\ref{tab:gconf}.
We use HYP staggered fermions as valence quarks which reduce the
taste-breaking for staggered fermions, and thus show improved chiral
behaviors \cite{Follana:2004sz, Follana:2005km,
  Follana:2006kb, Bae:2008qe}.
Meanwhile, the index theorem \cite{Atiyah:1963zz} states that 
\begin{align}
  Q_t &= n_- - n_+ \,,
  \qquad
  Q_t = \frac{1}{32\pi^2}\int_E d^4 x \;
  \epsilon_{\alpha\beta\mu\nu}\; \Tr(\, F_{\alpha\beta} \; F_{\mu\nu} )
  \label{eq:indexTheorem_cont}
\end{align}
where $Q_t$ is the topological charge \cite{ AliKhan:2001ym}, the
subscript ${}_E$ represents the Euclidean space, and $n_+$ ($n_-$) is
the number of zero modes with right-handed (left-handed) helicity.
In the continuum, Eq.~\eqref{eq:indexTheorem_cont} indeed comes from
the axial Ward identity \cite{Cundy:2016tmw}.
For staggered fermions, a similar relation holds but four-fold degeneracy
which comes from the approximate \SU{4} taste symmetry should be counted
\cite{Jeong:2017kst}:
\begin{equation}
  \label{eq:indexTheorem_stag}
  Q_t = \frac{1}{4} ( n^s_- - n^s_+ ) \,,
\end{equation}
where $n^s_{\pm}$ represent the number of zero modes with right-handed
($+$) and left-handed ($-$) helicities for staggered quarks.
Here, $n^s_{\pm}$ must be multiples of four due to the taste symmetry.

\section{Eigenvalue spectrum}
\label{sec:eigSpec}
\begin{figure}[t!]
  \captionsetup[subfigure]{aboveskip=-0.5em,belowskip=-0.5em}
  \centering
  \vspace*{-5mm}
  \begin{subfigure}{.49\linewidth}
    \centering
    \includegraphics[width=1.0\linewidth]{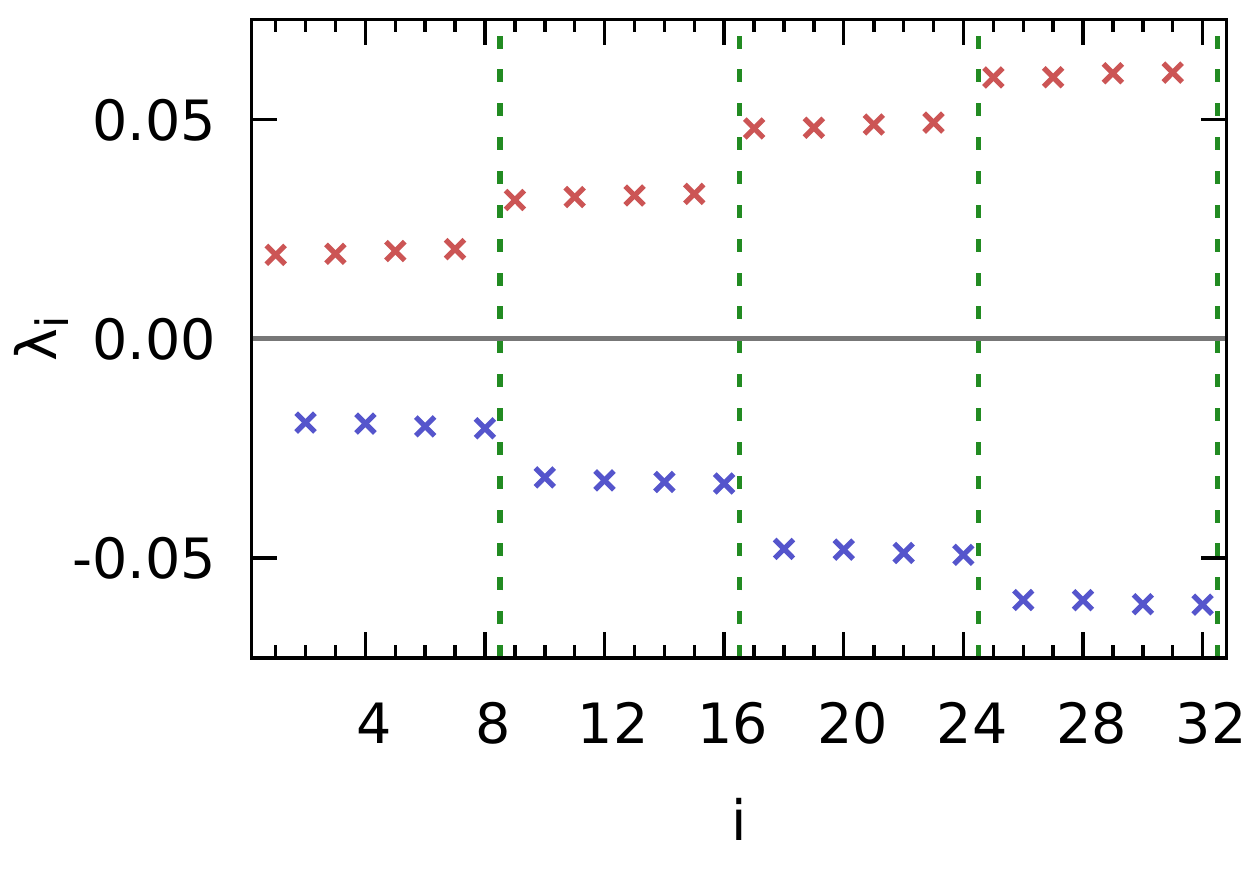}
    \caption{$Q_t=0$}
    \label{fig:eigVal_qt_0}
  \end{subfigure}
  \hspace{\fill}
  \begin{subfigure}{.49\linewidth}
    \centering
    \includegraphics[width=1.0\linewidth]{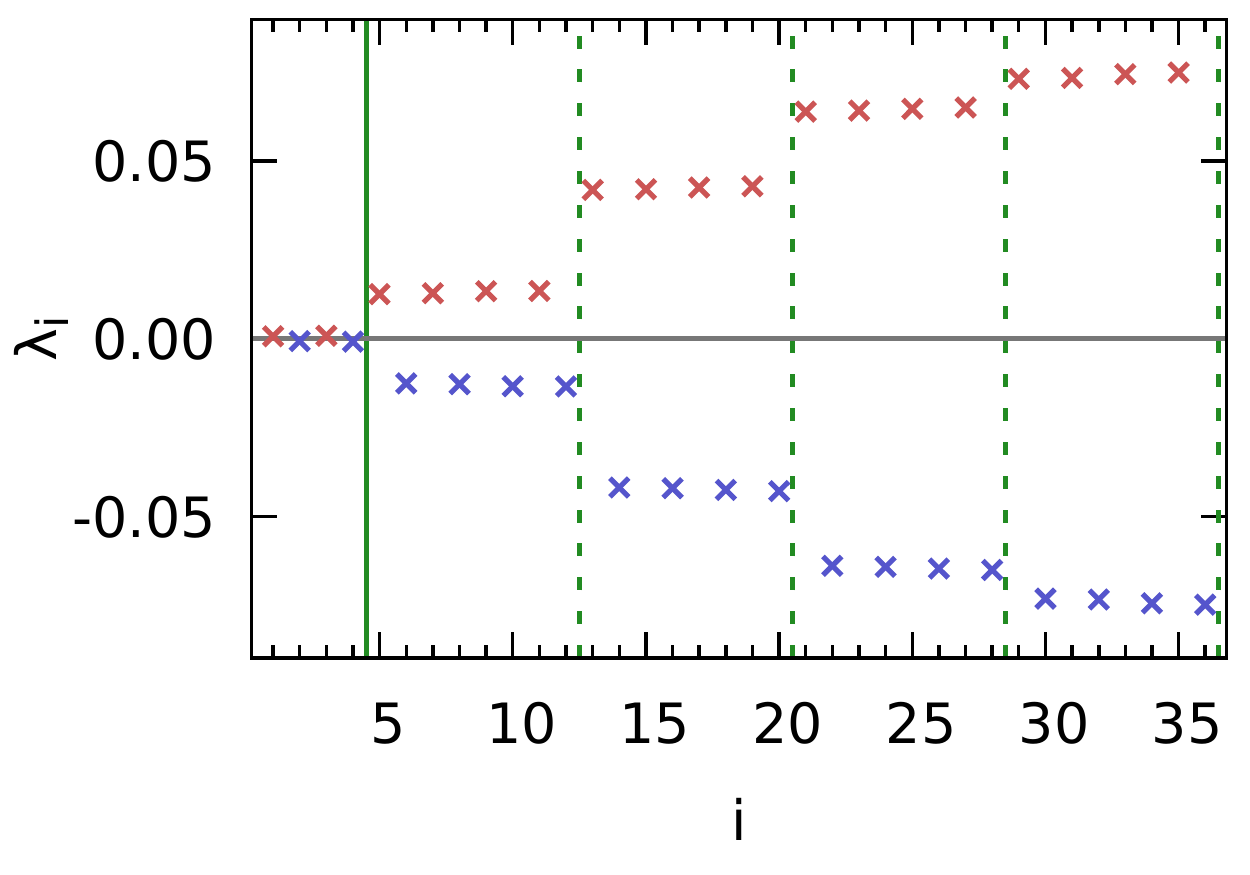}
    \caption{$Q_t=-1$}
    \label{fig:eigVal_qt_-1}
  \end{subfigure}
  \caption{Eigenvalue spectra of staggered Dirac operator on gauge
    configurations with $Q_t = 0$ and $Q_t = -1$. Here, $i$ represents an
    index of eigenvalue $\lambda_i$.}
  \label{fig:eigVal}
\end{figure}
In Fig.~\ref{fig:eigVal}, we present tens of low-lying eigenvalues of
the Dirac operator with HYP staggered quarks on gauge configurations
with topological charges $Q_t = 0, -1$.
Here, we measure $Q_t$ using the $Q(\text{5Li})$ operator defined in
Ref.~\cite{ deForcrand:1997esx, deForcrand:1995qq} after $10 \sim 30$
iterations of the APE smearing with $\alpha = 0.45$ \cite{ Cichy:2014qta,
  Hasenfratz:1998qk, Falcioni:1984ei}.
In the plot, eigenvalues are sorted in ascending order of their absolute
values $|\lambda_i|$.
Here, we assign the index ($2n$) of the eigenvalue such that it
satisfies $\lambda_{2n} = -\lambda_{2n-1}$.
Even the would-be zero modes have tiny but nonzero values of $\lambda_i$ at
finite lattice spacing $a \ne 0$.
Hence, each eigenvalue has its parity partner with opposite sign
even though it belongs to the would-be zero modes.
The solid green line for $Q_t = -1$ is drawn at the boarder between
the would-be zero modes and the nonzero modes.

For would-be zero modes, their eigenvalues are exactly zero in the
continuum limit, and so they are their own parity partners by
themselves.
The number of the would be zero modes must be multiple of four since
the \SU{4} taste symmetry is exactly conserved in the continuum.
In Fig.~\ref{fig:eigVal_qt_-1}, one can see the would-be zero modes appear
with four-fold degeneracy.
For nonzero modes, one eigenvalue must have four-fold degeneracy due to the
\SU{4} taste symmetry in the continuum, and its \UA{1} parity partner
should have the same four-fold degeneracy.
Hence, for each nonzero eigenvalue, it has a set of eight-fold degeneracy
due to the exact \UA{1} symmetry on the lattice and the \SU{4} taste
symmetry in the continuum.
In Fig.~\ref{fig:eigVal}, one can see the nonzero modes show up with
eight-fold degeneracy.

\section{Chirality for staggered fermions}
\label{sec:chiral}

Let us consider three chirality operators: $\Gamma_\varepsilon$,
$\Gamma_5$, and $\Xi_5$ defined as
\begin{align}
  \Gamma_\varepsilon & \equiv [ \gamma_5 \otimes \xi_5 ] \,,
  \qquad \Gamma_5  \equiv [ \gamma_5 \otimes \mathbb{1} ] \,,
  \qquad \Xi_5  \equiv [ \mathbb{1} \otimes \xi_5 ] \,.  
\end{align}
$\Gamma_\varepsilon$ represents a chirality of the conserved \UA{1}
symmetry for staggered fermions.
A taste singlet operator $\Gamma_5$ corresponds to the generator for
the anomalous \UAa{1} symmetry in the continuum.
Similarly, $\Xi_5$ represents the parity partner for the chirality
operator $\Gamma_5$.

The $\Gamma_\varepsilon$, $\Gamma_5$, and $\Xi_5$ operators satisfy
the same relations as the continuum chirality operator $\gamma_5$
as follows,
\begin{gather}
  (\Gamma)^{2n+1} = \Gamma \,,
  \qquad
  (\Gamma)^{2n} = \mathbb{1} \,,
  \label{eq:recursion2}
\end{gather}
where $\Gamma \in \{ \Gamma_\varepsilon, \Gamma_5, \Xi_5 \}$.
Furthermore, they are related to each other by
\begin{gather}
  \label{eq:epsilonDecomp}
  \Gamma_\varepsilon = \Gamma_5 \; \Xi_5 \;;\quad
  \Gamma_5 = \Xi_5 \; \Gamma_\varepsilon \;;\quad
  \Xi_5 = \Gamma_5 \; \Gamma_\varepsilon \,.
\end{gather}
These properties insure that they are the best choice to examine the
chiral symmetry for staggered fermions.
\begin{figure}[t!]
  \captionsetup[subfigure]{aboveskip=-1.5em,belowskip=-1.0em}
  \captionsetup{aboveskip=-0em,belowskip=0em}
  \centering
  \vspace*{-1.5em}
  \begin{subfigure}[t]{.24\linewidth}
    \centering
    \includegraphics[width=1.0\linewidth]{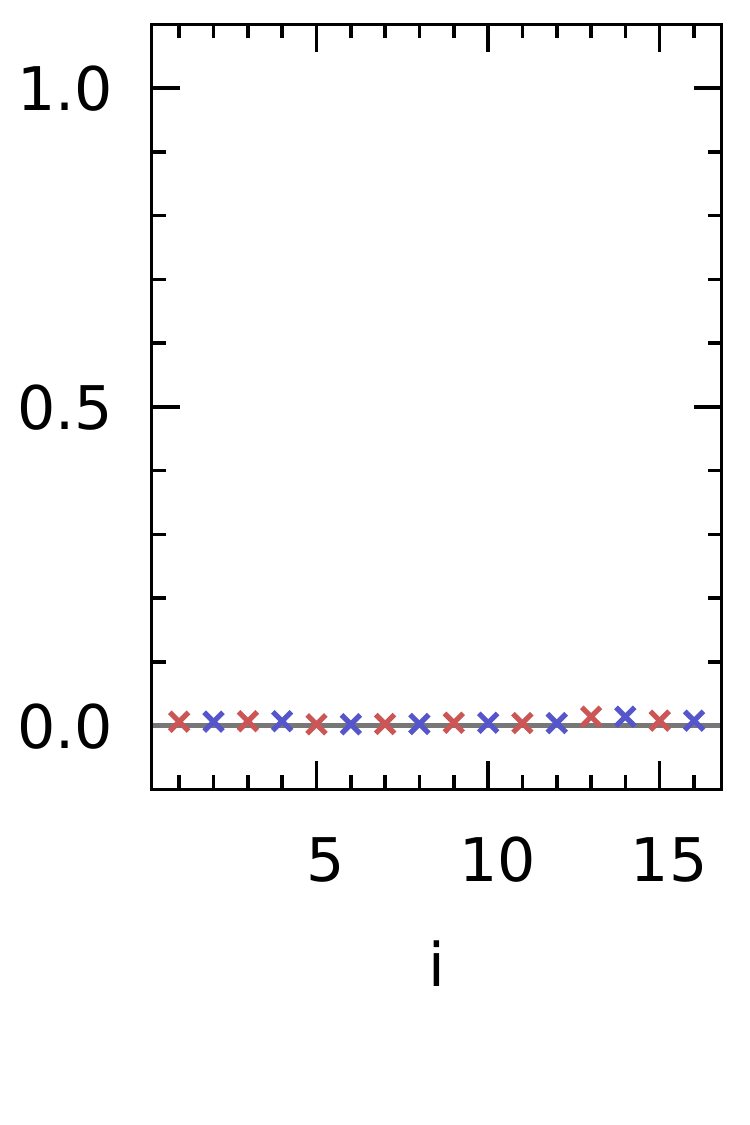}
    \caption{$Q_t=0$}
    \label{fig:chiral_qt_0}
  \end{subfigure}
  %
  %
  \begin{subfigure}[t]{.24\linewidth}
    \centering
    \includegraphics[width=1.0\linewidth]{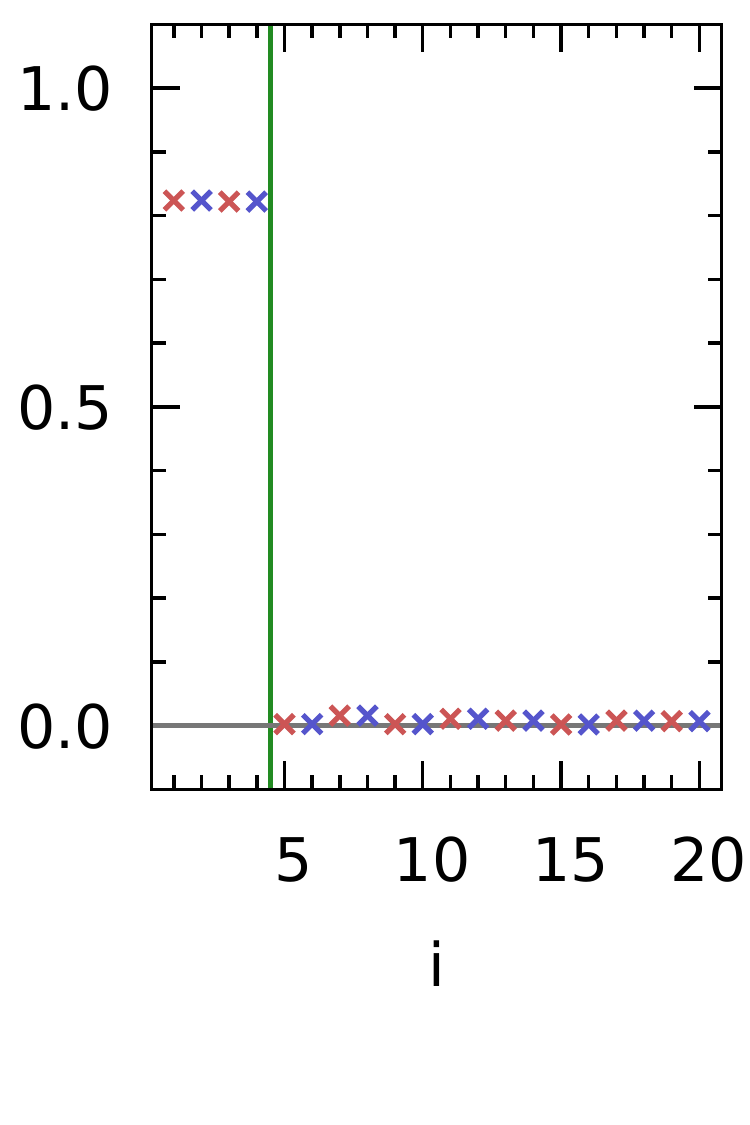}
    \caption{$Q_t=-1$}
    \label{fig:chiral_qt_-1}
  \end{subfigure}
  \begin{subfigure}[t]{.24\linewidth}
    \centering
    \includegraphics[width=1.0\linewidth]{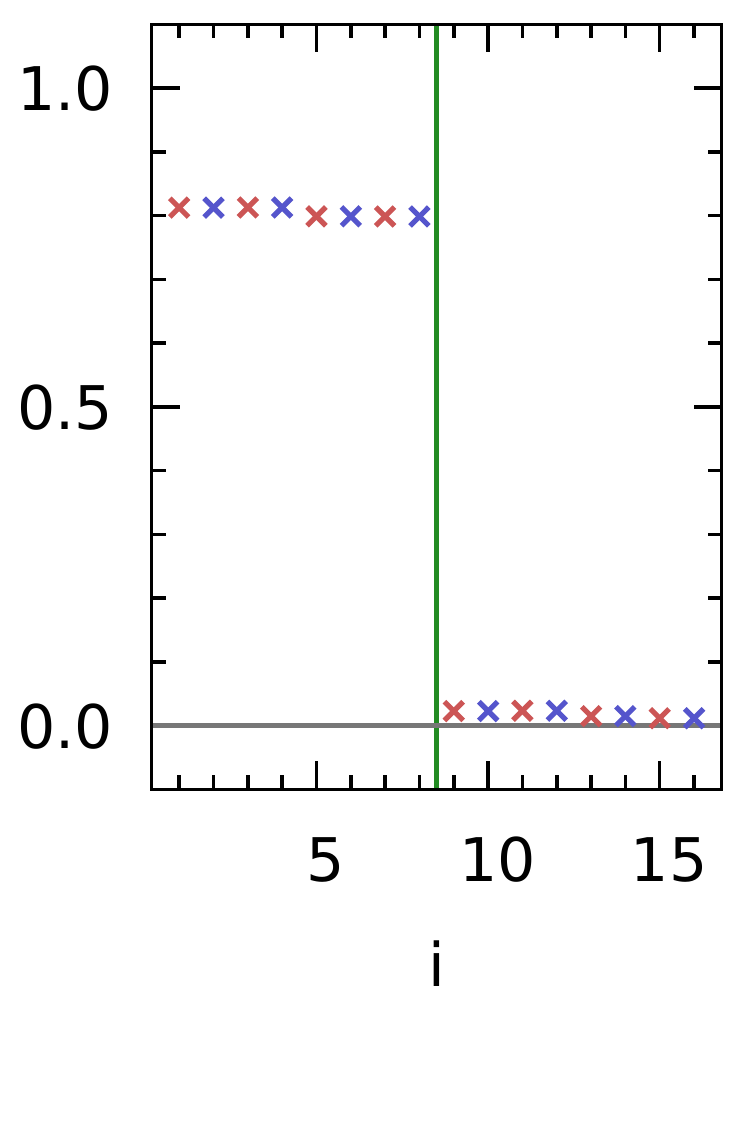}
    \caption{$Q_t=-2$}
    \label{fig:chiral_qt_-2}
  \end{subfigure}
  %
  %
  \begin{subfigure}[t]{.24\linewidth}
    \centering
    \includegraphics[width=1.0\linewidth]{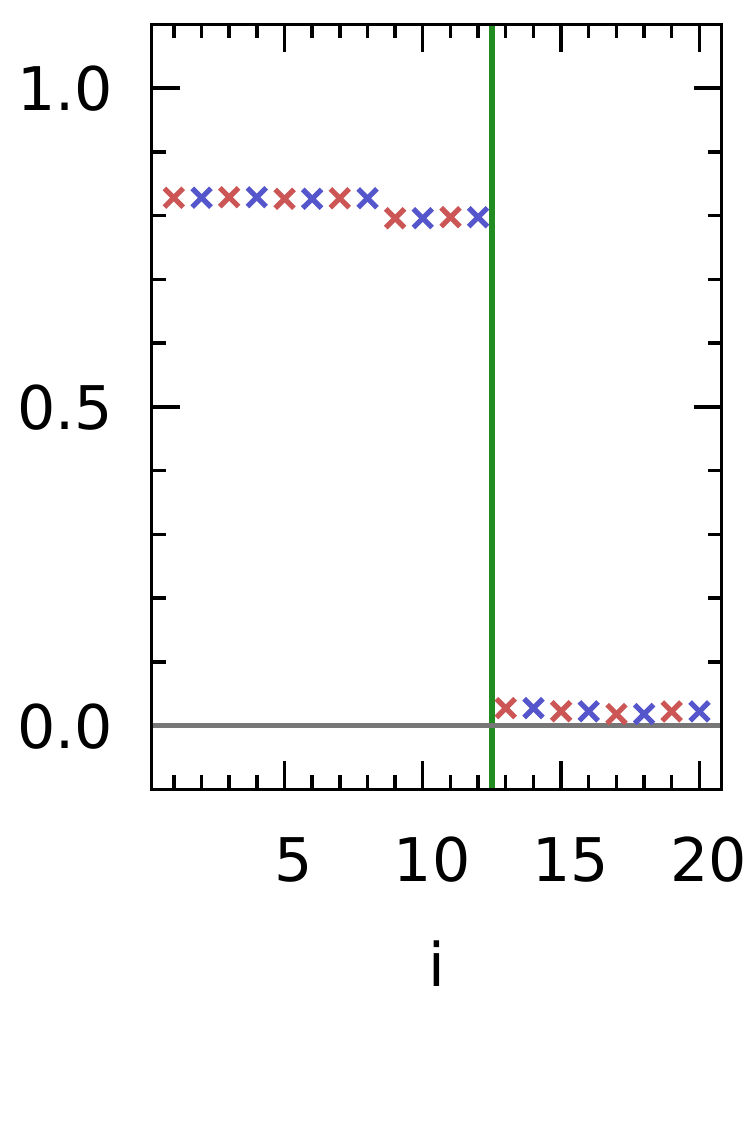}
    \caption{$Q_t=-3$}
    \label{fig:chiral_qt_-3}
  \end{subfigure}
  \caption{$\Gamma_5(\lambda_i)$ for various topological charges. Here, $i$
    represents an index of eigenvalue $\lambda_i$.}
  \label{fig:chiral}
\end{figure}
Let us define the chirality as
\begin{align}
  \Gamma_5(\alpha,\beta) &\equiv
  \langle f_{\alpha} | [ \gamma_5 \otimes \mathbb{1} ] | f_{\beta} \rangle\,,
  \qquad
  \Gamma_5(\lambda_i) \equiv \Gamma_5(\lambda_i,\lambda_i) \,.
\end{align}
In Fig.~\ref{fig:chiral}, we measure the chirality $\Gamma_5(\lambda_i)$ for
topological charges $Q_t = 0, -1, -2, -3$, respectively.
Comparing with Fig.~\ref{fig:eigVal}, the would-be zero modes has a
non-trivial chirality around 0.8 in magnitude, while nonzero modes
have values of $\Gamma_5(\lambda_i)$ close to zero.
Consequently, would-be zero modes are manifestly distinguishable from
nonzero modes by the $\Gamma_5$ chirality as shown in Ref.~\cite{
  Follana:2004sz, Follana:2005km, Follana:2006kb}.
The magnitudes of the chirality for would-be zero modes are somewhat
smaller than one, the continuum expectation value.
It is because the $\Gamma_5$ operator is not conserved at $a\ne 0$ and
receives a finite renormalization on the lattice.
%

%



\section{Chiral Ward identity}
\label{sec:ward}
Rewriting Eqs.~\eqref{eq:fp_to_fm} by implementing
Eq.~\eqref{eq:recursion2} and Eq.~\eqref{eq:epsilonDecomp}, we obtain the
following chiral Ward identities for staggered fermions:
\begin{align}
  \label{eq:ward1}
  \Gamma_5 \, | f_{+\lambda} \rangle &= e^{+i\theta} \, \Xi_5 \, |
  f_{-\lambda} \rangle \,,
  \qquad
  \Gamma_5 \, | f_{-\lambda} \rangle = e^{-i\theta} \, \Xi_5 \, |
  f_{+\lambda} \rangle \,.
\end{align}
Let us define the chirality matrix elements sandwiched between the
two eigenvectors as
\begin{align}
  \label{eq:matEpsilon}
  \Gamma_\varepsilon ( \alpha, \beta )
  &\equiv \langle f_\alpha | \Gamma_\varepsilon | f_\beta \rangle
  = \langle f_\alpha |[ \gamma_5 \otimes \xi_5 ] | f_\beta \rangle \,,
  \\
  \label{eq:matGamma5}
  \Gamma_5 ( \alpha, \beta )
  &\equiv \langle f_\alpha | \Gamma_5 | f_\beta \rangle
  = \langle f_\alpha |[ \gamma_5 \otimes \mathbb{1} ] | f_\beta \rangle
  \,,
  \\
  \label{eq:matXi5}
  \Xi_5 ( \alpha, \beta )
  &\equiv \langle f_\alpha | \Xi_5 | f_\beta \rangle
  = \langle f_\alpha |[ \mathbb{1} \otimes \xi_5 ] | f_\beta \rangle
  \,.
\end{align}
Using the Ward identity of Eqs.~\eqref{eq:ward1}, we rewrite the chirality
matrix elements as follows,
\begin{align}
  \Gamma_5 (\alpha,+\beta)
  & = e^{+i\theta_\beta} \; \Xi_5 (\alpha,-\beta) \,,
  \qquad
  \Gamma_5 (\alpha,-\beta)
  = e^{-i\theta_\beta} \; \Xi_5 (\alpha,+\beta) \,,\\
  \Gamma_5 (+\alpha,\beta)
  & = e^{-i\theta_\alpha} \; \Xi_5 (-\alpha,\beta) \,,
  \qquad
  \Gamma_5 (-\alpha,\beta)
  = e^{+i\theta_\alpha} \; \Xi_5 (+\alpha,\beta) \,.
\end{align}
If we take the norm of them, then 
\begin{equation}
  | \; \Gamma_5 (\alpha,\beta) \, |
  = | \; \Xi_5 (\alpha,-\beta) \, |
  = | \; \Xi_5 (-\alpha,\beta) \, |
  = | \; \Gamma_5 (-\alpha,-\beta) \, | \,.
\end{equation}
In addition, the Hermiticity insures interchanging $\alpha$ and
$\beta$, which provides the final form of the Ward identities:
\begin{align}
  | \; \Gamma_5 (\alpha,\beta) \, |
  & = | \; \Xi_5 (\alpha,-\beta) \, |
  = | \; \Xi_5 (-\alpha,\beta) \, |
  = | \; \Gamma_5 (-\alpha,-\beta) \, | \nonumber \\
  & = | \; \Gamma_5 (\beta,\alpha) \, |
  = | \; \Xi_5 (\beta,-\alpha) \, |
  = | \; \Xi_5 (-\beta,\alpha) \, |
  = | \; \Gamma_5 (-\beta,-\alpha) \, | \,.
  \label{eq:ward3}
\end{align}
\begin{table}[t!]
  \captionsetup{aboveskip=0.5em}
  \small
  \renewcommand{\arraystretch}{1.2}
  \vspace*{-1.5em}
  \begin{subtable}{.3\linewidth}
    \begin{tabular}[b]{ >{\centering\arraybackslash}m{0.42\linewidth}
          >{\centering\arraybackslash}m{0.38\linewidth} }
      \hline\hline
      parameter & value \\
      \hline
      $|\,\Gamma_5 (\lambda_1,\lambda_1)\,|$ & 0.8238257 \\
      $|\,\Xi_5 (\lambda_2,\lambda_1)\,|$ & 0.8238257 \\
      $|\,\Xi_5 (\lambda_1,\lambda_2)\,|$ & 0.8238257 \\
      $|\,\Gamma_5 (\lambda_2,\lambda_2)\,|$ & 0.8238257 \\
      \hline\hline
    \end{tabular}
    \caption{Diagonal WI}
    \label{tab:wardDiag}
  \end{subtable}
  %
  \qquad
  \begin{subtable}{.6\linewidth}
    \begin{tabular}{ >{\centering\arraybackslash}m{0.25\linewidth}
        >{\centering\arraybackslash}m{0.20\linewidth}
        | >{\centering\arraybackslash}m{0.25\linewidth}
        >{\centering\arraybackslash}m{0.20\linewidth}
        }
      \hline\hline
      parameter & value & parameter & value \\
      \hline
      $|\,\Gamma_5 (\lambda_{75}, \lambda_{70})\,|$ & 0.5008622 &
      $|\,\Gamma_5 (\lambda_{70}, \lambda_{75})\,|$ & 0.5008622 \\
      $|\,\Xi_5 (\lambda_{69}, \lambda_{75})\,|$ & 0.5008622 &
      $|\,\Xi_5 (\lambda_{75}, \lambda_{69})\,|$ & 0.5008622 \\
      $|\,\Xi_5 (\lambda_{70}, \lambda_{76})\,|$ & 0.5008622 &
      $|\,\Xi_5 (\lambda_{76}, \lambda_{70})\,|$ & 0.5008622 \\
      $|\,\Gamma_5 (\lambda_{69},\lambda_{76})\,|$ & 0.5008622 &
      $|\,\Gamma_5 (\lambda_{76},\lambda_{69})\,|$ & 0.5008622 \\
      \hline\hline
    \end{tabular}
    \caption{Off-diagonal WI}
    \label{tab:wardOffdiag}
  \end{subtable}
  \caption{Numerical demonstration of chiral Ward identity (WI) in
    Eq.~\eqref{eq:ward3}. Here, $\lambda_2 = -\lambda_1$,
    $\lambda_{70} = -\lambda_{69}$, $\lambda_{76} = -\lambda_{75}$.}
  \label{tab:ward}
\end{table}
Table~\ref{tab:ward} shows how the chiral Ward identities of
Eq.~\eqref{eq:ward3} works in our numerical study.
Here, it confirms that they are valid within our numerical precision.
%




\section{Leakage of chirality}
\label{sec:leak}
Here, we focus on off-diagonal elements ($\alpha \neq \beta$) of
chirality $\Gamma_5 ( \alpha, \beta )$ and $\Xi_5 ( \alpha, \beta
)$.
We are interested in how much of the chirality of an eigenmode leaks
into other eigenmodes.
Traditionally, the diagonal elements of chirality $\Gamma_5(\lambda_i)$
were measured and studied as in Ref.~\cite{ Follana:2004sz, Follana:2005km,
  Follana:2006kb}.
Here, we study on the off-diagonal elements of chirality
$\Gamma_5(\alpha,\beta)$, and $\Xi_5(\alpha,\beta)$ with
$\alpha \ne \beta$.
In the continuum, the \SU{4} taste symmetry is respected.
The net consequence of the conserved taste symmetry is that each
nonzero eigenvalue has eight-fold degeneracy, and these eight
degenerate eigenmodes will mix with one another within the eight-fold
degenerate members.
In other words, if $|\alpha| \ne |\beta|$, then
  $\Gamma_5(\alpha,\beta) = \Xi_5(\alpha,\beta) = 0$ in the continuum
($a=0$) thanks to the \SU{4} taste symmetry.
However, at finite lattice ($a\ne 0$), the \SU{4} taste symmetry is
not exact, but mostly respected near the continuum.
Hence, the leakage from one set of the eight-fold degeneracy to other
set of eight-fold degeneracy will be very small near the continuum
($a \approx 0$).
Therefore, it will be very interesting to study this leakage pattern
in the chirality measurement.
\begin{figure}[t!]
  \centering
  \captionsetup[subfigure]{aboveskip=-0.25em,belowskip=-0.5em}
  \vspace*{-2.5em}
  \begin{subfigure}{.49\linewidth}
    \centering
    \includegraphics[width=1.0\linewidth]
                    {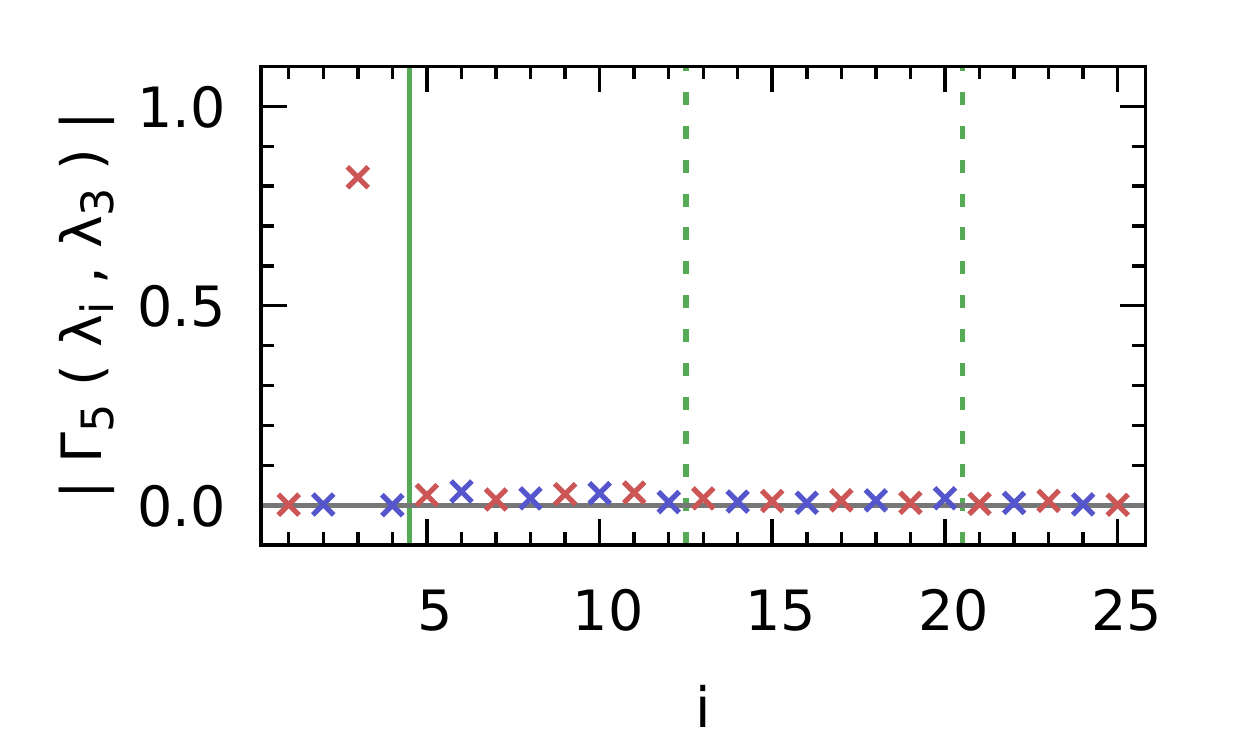}
    \caption{$Q_t = -1$}
    \label{fig:leak_g5x1_f3}
  \end{subfigure}
  \hfill
  \begin{subfigure}{.49\linewidth}
    \centering
    \includegraphics[width=1.0\linewidth]
                    {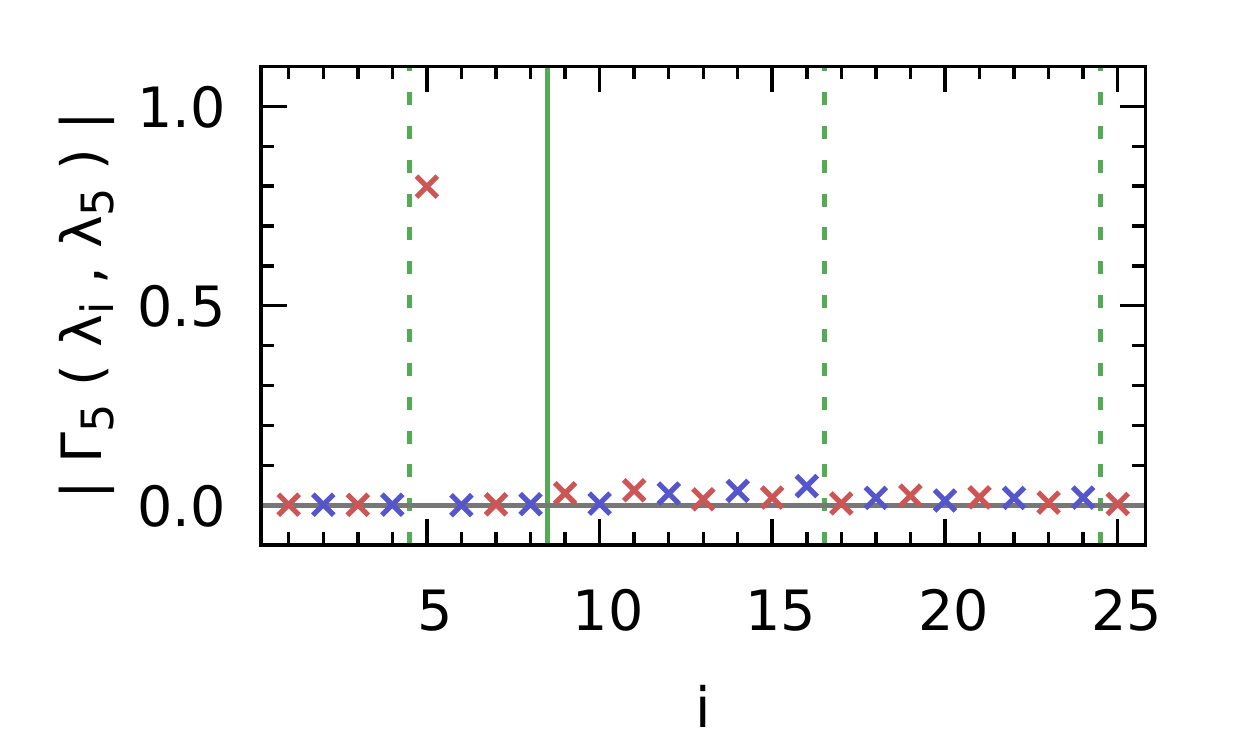}
    \caption{$Q_t = -2$}
    \label{fig:leak_1xxi5_f3}
  \end{subfigure}
  \caption{Leakage patterns of would-be zero modes for the $\Gamma_5$
    operator.}
  \label{fig:leakage_zero}
\end{figure}
Fig.~\ref{fig:leakage_zero} shows leakage patterns of $\Gamma_5$ for
would-be zero modes.
Here, we observe that for $\Gamma_5$ there is only one non-trivial
signal at the would-be zero mode itself and there is almost no leakage
to nearest zero and nonzero eigenmodes.
\begin{figure}[b!]
  \centering
  \captionsetup[subfigure]{aboveskip=-0.25em,belowskip=-0.5em}
  \captionsetup{aboveskip=1em,belowskip=-1.0em}
  \vspace{-2.0em}
  \begin{subfigure}{.49\linewidth}
    \centering
    \includegraphics[width=1.0\linewidth]{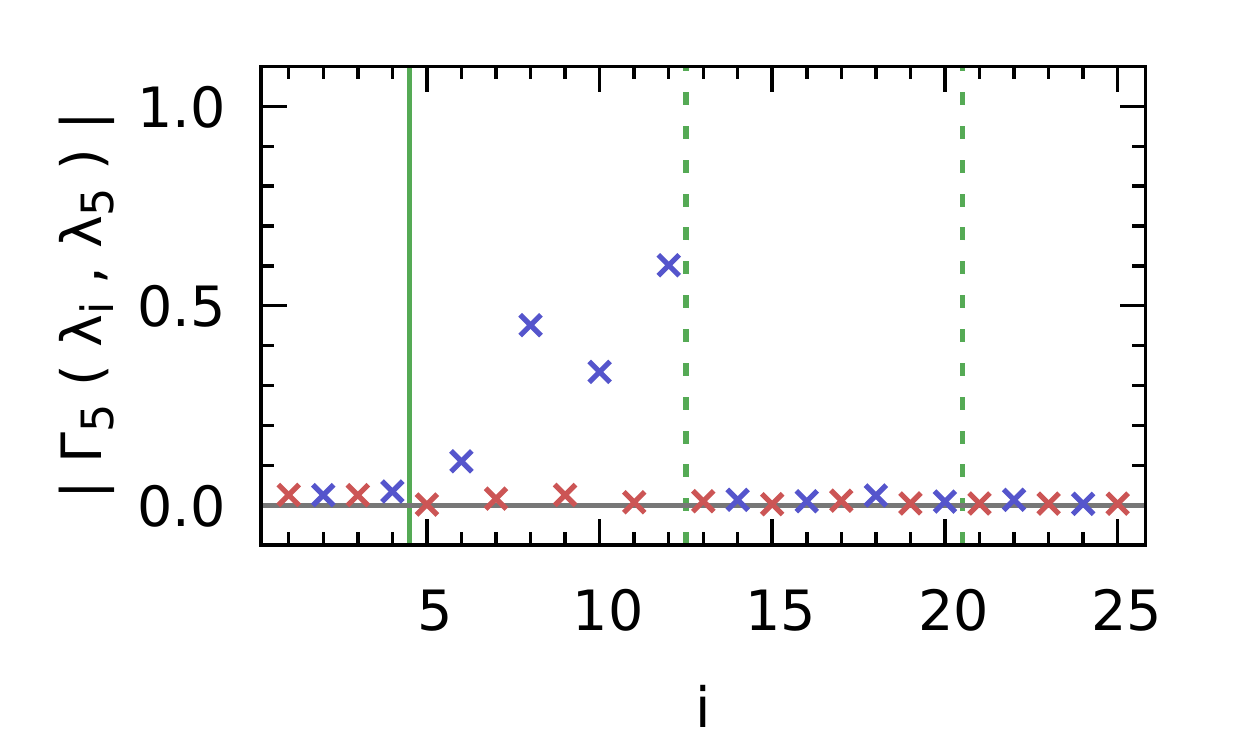}  
    \caption{Leakage for $\lambda_5$ with $Q_t=-1$}
    \label{fig:leak_g5x1_f5}
  \end{subfigure}
  \hfill
  \begin{subfigure}{.49\linewidth}
    \centering
    \includegraphics[width=1.0\linewidth]{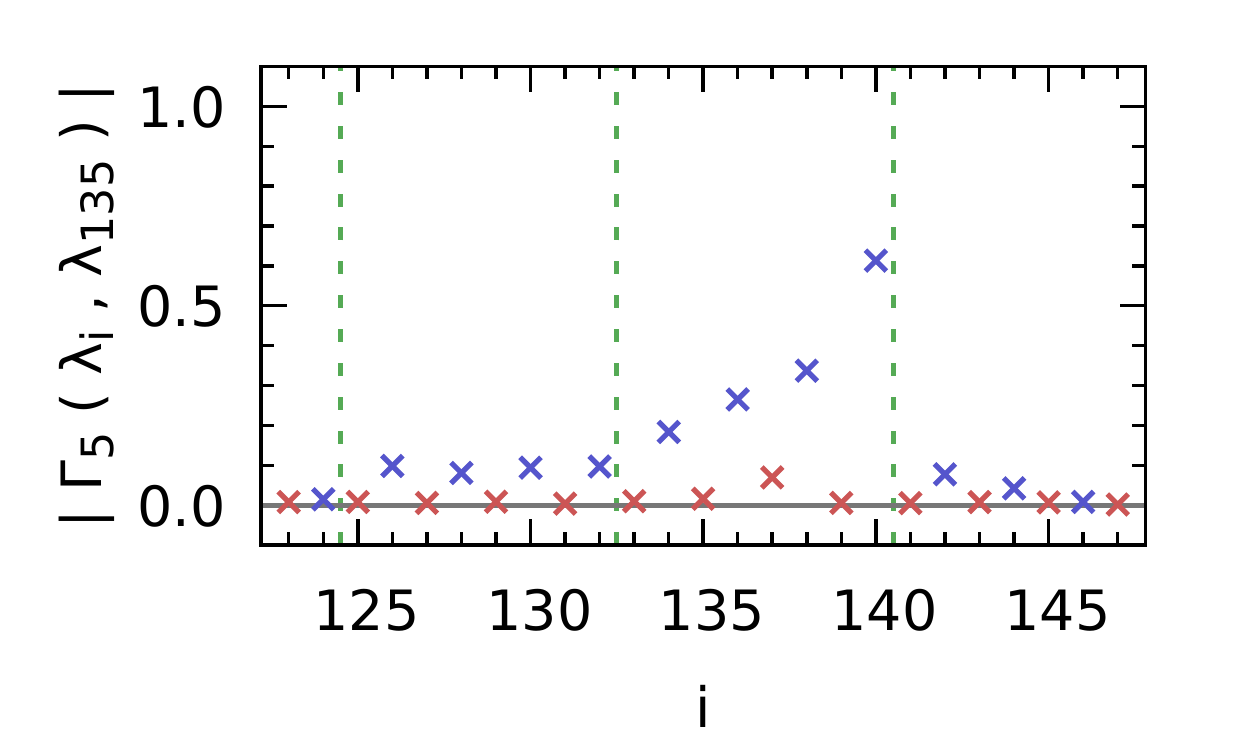}  
    \caption{Leakage for $\lambda_{135}$ with $Q_t=-1$}
    \label{fig:leak_1xxi5_f135}
  \end{subfigure}
  \caption{Leakage patterns of $\Gamma_5$ for nonzero eigenmodes.}
  \label{fig:leakage_nonzero}
\end{figure}
For nonzero modes, leakage of the $\Gamma_5$ operator for an
eigenvalue $\lambda_i$ is supposed to go into the four-fold parity
partners with eigenvalue $-\lambda_i$ in the continuum.
Near the continuum ($a \approx 0$), the \SU{4} taste symmetry is almost
respected and the leakage to eigenmodes outside the eight-fold
degeneracy is also almost prohibited.
This kind of leakage patterns for nonzero eigenmodes are presented in
Fig.~\ref{fig:leakage_nonzero}.
In Fig.~\ref{fig:leak_g5x1_f5}, the leakage of the $\Gamma_5$ chirality for
the eigenmode $| f_{\lambda_5} \rangle$ goes into the eigenmodes
with eigenvalue $-\lambda_5$: $| f_{\lambda_j} \rangle$ with $j = 6,
8, 10, 12$, as the theory predicts.
In Fig.~\ref{fig:leak_1xxi5_f135}, the leakage of the $\Gamma_5$ chirality
for the eigenmode $| f_{\lambda_{135}} \rangle$ goes into the eigenmodes
with eigenvalue $-\lambda_{135}$: $| f_{\lambda_j} \rangle$ with $j = 134,
136, 138, 140$.
Here, we also observe a small effect of the \SU{4} taste symmetry breaking
in that a small amount of leakage of the $\Gamma_5$ chirality for the
eigenmode $| f_{\lambda_{135}} \rangle$ goes into eigenmodes outside of the
eight-fold degeneracy such as $| f_{\lambda_j} \rangle$ with $j=126, 128,
130, 132, 142, 144$.
%




\section{Conclusion}
\label{sec:conc}
We have studied $\Gamma_5$ and $\Xi_5$ chirality for eigenmodes of
staggered fermions.
Thanks to the Ward identities, $\Gamma_5$ chirality is completely
correlated with $\Xi_5$ chirality.
We demonstrate how the leakage patterns of $\Gamma_5$ chirality can
be used to distinguish zero eigenmodes and nonzero eigenmodes.
%

\acknowledgments{
We would like to express our sincere gratitude to Eduardo Follana for his
kind help.
The research of W.~Lee is supported by the Mid-Career Research Program
(Grant No.~NRF-2019R1A2C2085685) of the NRF grant funded by the Korean
government (MOE).
This work was supported by Seoul National University Research Grant in
2019.
W.~Lee would like to acknowledge the support from the KISTI supercomputing
center through the strategic support program for the supercomputing
application research (No.~KSC-2017-G2-0009).
Computations were carried out on the DAVID cluster at Seoul National
University.
}

\bibliographystyle{JHEP}

\bibliography{ref}

\end{document}